\documentclass{an}
\usepackage{graphicx}
\usepackage{times}
\usepackage{fancyhdr}
\usepackage{natbib}

\begin{document}

\title{Boundary layer emission in luminous LMXBs}   

\author{M.Gilfanov\inst{1,2}
\and
M.Revnivtsev\inst{1,2}
}
\institute{Max-Planck-Institute f\"ur Astrophysik,
Karl-Schwarzschild-Str. 1, D-85740 Garching bei M\"unchen,
Germany,
\and
Space Research Institute, Russian Academy of Sciences,
Profsoyuznaya 84/32, 117997 Moscow, Russia
            }
\date{}

   \abstract{We show that aperiodic and quasiperiodic variability of
bright LMXBs -- atoll and Z- sources,  on $\sim$ sec -- msec time
scales  is caused primarily by variations of the luminosity of the
boundary layer. 
The emission of the accretion disk is  less variable on
these time scales and its power density spectrum follows $P_{\rm
disk}(f)\propto f^{-1}$  law, contributing to observed flux
variation at low frequencies and low energies only. 
The kHz QPOs have the same origin as variability at lower frequencies, 
i.e. independent of the nature of the "clock", the actual luminosity 
modulation takes place on the neutron star surface.
The boundary layer spectrum remains nearly constant in the course of 
the luminosity variations  and is represented to certain accuracy by
the  Fourier frequency resolved spectrum. In the investigated
range of $\dot{M}\sim (0.1-1) \dot{M}_{\rm Edd}$ it depends weakly on
the global mass accretion rate and in the limit $\dot{M}\sim
\dot{M}_{\rm Edd}$ is close to Wien spectrum with $kT\sim
2.4$ keV.
Its independence on the global value of $\dot{M}$ lends support to
the  theoretical suggestion by \citet{inogamov99} that the  boundary
layer is radiation pressure supported. \\
Based on the knowledge of the boundary layer spectrum we attempt
to relate the motion along the Z-track to changes of physically
meaningful parameters. Our results suggest that the contribution of
the boundary layer to the observed emission decreases along the
Z-track from conventional $\sim 50\%$ on the horizontal branch to a
rather small number on the normal branch. This decrease can be
caused, for example, by obscuration of the boundary layer by the
geometrically thickened accretion  disk at 
$\dot{M}\sim\dot{M}_{\rm Edd}$. 
Alternatively, this can indicate significant change of the
structure of the accretion flow at $\dot{M}\sim\dot{M}_{\rm Edd}$
and disappearance of the boundary layer as a distinct region of the 
significant energy release associated with the neutron star surface. 
   }

   \maketitle

%
%________________________________________________________________

\section{Introduction}
\label{sec:intro}

Accreting neutron stars in low mass X-ray binaries (LMXB) are
among the most luminous compact X-ray sources in the Milky Way. 
A number of them have luminosities exceeding $\sim {\rm
few}\times 10^{38}$ erg/s and presumably  accrete matter at the level
close to the critical Eddington accretion rate. 
In the bright state these sources have
rather soft X-ray spectra, indicating that their X-ray
emission is predominantly formed in the optically thick media. 
Similar to black holes,  at lower luminosities, 
$\log(L_{\rm X})\la 36.5-37$, 
neutron stars undergo a transition to the hard spectral state
\citep[e.g.][]{barret01}. The energy spectra in this state
point at the low optical depth in the emission region, indicating a
significant change of the geometry of the accretion flow.

In the soft spectral state, the commonly accepted picture of accretion
at not too extreme values of  accretion rate has
two main ingredients -- the  accretion disk (AD) and the boundary
layer (BL).  
While in the disk matter rotates with nearly Keplerian velocities, in
the boundary layer it decelerates down to the spin frequency of the
neutron star and settles onto its surface. 
For the typical neutron star spin frequency ($\la500-700$Hz) comparable
amounts of energy are released in  these two regions
\citep{ss86,sibg00}.  
This picture is based on rather obvious qualitative expectations as
well as more sophisticated theoretical considerations and numerical
modeling \citep{ss86,kluzniak,inogamov99,sibg00}.
It has been receiving, however, little direct observational
confirmation.  Due to similarity of the spectra of the accretion disk
and boundary layer the total spectrum has a smooth curved shape, which
is difficult to decompose into separate spectral components
\citep{mitsuda84,white88,disalvo01,done02}.   
This made  application of physically motivated spectral models to the
description of observed spectra of luminous neutron stars difficult,
in spite of very significant increase in the sensitivity of 
X-ray instruments. Not surprisingly, the best fit parameters derived
from the data of different instruments and, correspondingly, the
inferred values of the physically meaningful quantities  are often in
contradiction to each other.  

This ambiguity can be resolved if the spectral information is analysed
together with  timing data.
Early results of \citet{mitsuda84} and \citet{mitsuda86} suggested
that the boundary layer and accretion disk may have different patterns
of spectral variability.
Based on the TENMA data, they studied the difference between the
spectra averaged at different intensity levels -- that restricted the
range of accessible time scales to $\ga 10^3$ sec.
\cite{gilfanov03} and \cite{mikej05} have exploited the technique of
Fourier frequency resolved spectroscopy \citep{freq_res99} to study
spectral variability of luminous LMXBs in a broad range of time
scales, including kHz QPO.  Their findings are reviewed and discussed 
below.

\begin{figure}
\includegraphics[width=0.5\textwidth]{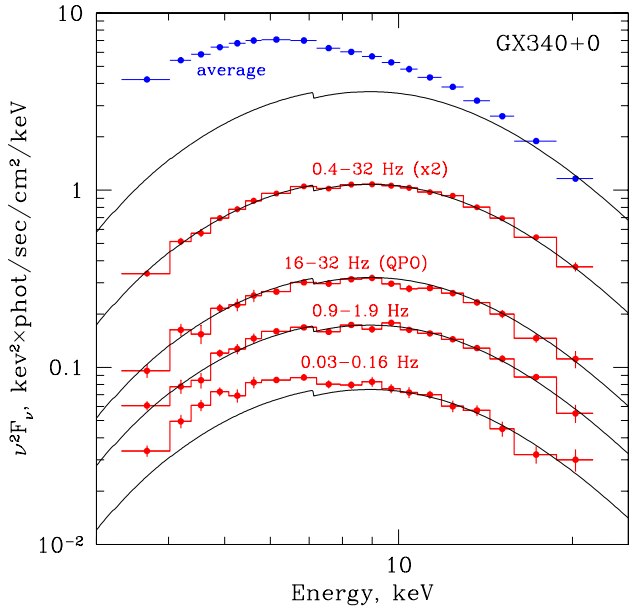}
\caption{Average and frequency resolved spectra of GX340+0 on the
horizontal branch of the color-color diagram. The solid lines show the  
Comptonization  spectrum with parameters similar to those given in the
Section \ref{sec:bl_spectrum}. 
\label{fig:freqres_gx340}}
\end{figure}

\begin{figure}
\includegraphics[width=0.5\textwidth]{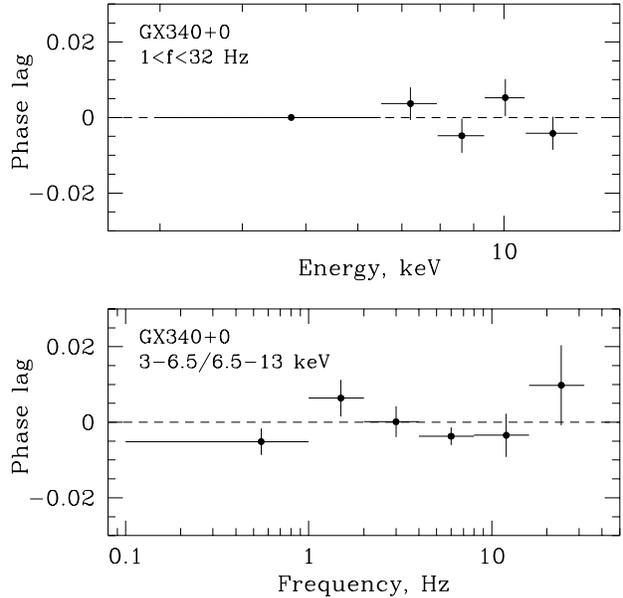}
\caption{Phase lags in GX340+0 on the horizontal branch as function of
energy ({\em upper panel}) and Fourier frequency ({\em lower
panel}). The energy dependent phase 
lags were computed in the 1--32 Hz frequency range, the frequency
dependent lags are between 3--6.5 keV and 6.5--13 keV energy
bands. The phase is normalized to 0--1 interval.
\label{fig:lags_gx340}}
\end{figure}

\section{Fourier-frequency resolved spectroscopy of luminous LMXBs}
\label{sec:freqres_theory}

\subsection{The method}

As defined in \citet{freq_res99}, the Fourier frequency resolved spectrum
is the energy dependent rms amplitude in a selected frequency range,
expressed in absolute  (as opposite to fractional) units. 
A similar approach was used by \citet{mendez0614} to study the energy
spectrum of kHz oscillations in  4U0614+09.
One of its advantages over fractional rms--vs.--energy dependence is
the possibility to use conventional (i.e. response folded) spectral
approximations in order to describe the energy dependence of aperiodic
variability. Although the interpretation of the frequency resolved
spectra is not always straightforward, several applications of this
technique to variability of black hole binaries gave meaningful
results \citep[e.g.][]{freq_res99, gilfanov_cygx1}.

We will use the following example as an illustration.
Let us consider a two-component spectrum, consisting of a constant and
variable component. The variable component changes its normalization
but not the spectral shape.
In this case the shape of the frequency resolved
spectrum would not depend on the Fourier frequency and would be
identical to the spectrum of the variable component.  
Importantly, the X-ray flux in all energy 
channels  will vary coherently and with zero time/phase lag between
different energies. Presence of significant phase lag  and/or Fourier
frequency dependence of the frequency resolved spectra would indicate
that a more complex pattern of spectral  variability is taking place.

With few exceptions \citep{dieters00}, phase lag between light curves
in different energy bands in luminous LMXBs 
is usually small, $\Delta\phi\la{\rm few}\times 10^{-2}$, 
coherence is consistent with unity,  \citep[e.g.][]{vaughan94, 
vaughan99, dieters00} and the behavior of the fractional
rms-vs-energy dependence is similar at different Fourier frequencies
\citep{vdk86, vdk00}. This suggests, that Fourier frequency resolved  
spectra can be interpreted in a  straightforward and model-independent
manner.

\subsection{Results}

For the case study we use archival data of PCA observations of a
Z-source GX340-0 on the horizontal branch of the color-color diagram.
\citet{gilfanov03} conducted similar study of an atoll source
4U1608-52 and arrived at similar conclusions. 
Our choice was defined by the requirement that the PCA configuration
combined sufficient energy resolution (large number of the energy
channels)  with good timing resolution and large total exposure  
time.  
The Fourier frequency resolved spectra in several frequency bands
corresponding to the band limited continuum noise component and  the
$\sim 25$ Hz QPO are shown in Fig.~\ref{fig:freqres_gx340} 
along with the conventional spectrum of the source averaged over the same
data.
The figure clearly demonstrates that shape of the spectra depends on
the Fourier frequency at low frequencies and becomes independent of
the frequency at  $f\ga 0.5$ Hz.  
Another conclusion from
the data presented in Fig.~\ref{fig:freqres_gx340}, important for the
following discussion,  is that  all frequency resolved spectra are
significantly harder than the average source spectrum.

The phase lags as function of energy and Fourier frequency are shown
in Fig.~\ref{fig:lags_gx340}. No statistically significant phase lags
were detected  with an upper limit of $\Delta\phi\sim 10^{-2}$, where
phase $\phi$ is normalized to the interval 0--1 (as opposed to
$0-2\pi$).

\subsection{Interpretation}

We show below that independence of the frequency
resolved spectra on the Fourier frequency and the smallness of the phase
lags require a particularly simple form of the spectral variability. 

The constancy of the spectral shape with Fourier frequency implies that the
power spectrum $P(E,\omega)$ can be represented as a product of two
functions,  one of which depends on the energy and the other on the
frequency only. For convenience we write $P(E,\omega)$ in the form:  
\begin{equation} 
P(E,\omega)=S^2(E)\times f^2(\omega)
\label{eq:pds}
\end{equation} 
where non-negative functions $S(E)$ and $f(\omega)$ can be
directly determined from the frequency resolved spectra.
The Fourier image of the light curve $F(E,t)$ is:
\begin{equation} 
\hat F(E,\omega)=S(E)\times f(\omega)\times e^{i \phi(E,\omega)}
\end{equation}
In the general case the complex argument $\phi(E,\omega)$ can depend both
on Fourier frequency $\omega$ and energy $E$. 
If the phase lags between different energies are negligibly small,
$\phi$ depends on the Fourier frequency only and the 
Fourier image of $F(E,t)$ is: 
\begin{equation} 
\hat F(E,\omega)=S(E)\times f(\omega)\times e^{i \phi(\omega)}
\label{eq:ft}
\end{equation}
The light curve $F(E,t)$ can be computed via inverse Fourier
transform of $\hat F(E,\omega)$:
\begin{eqnarray}
F(E,t)=\int d\omega \hat F(E,\omega) e^{i\omega t}=\nonumber\\
=S(E)\times \int d\omega f(\omega) e^{i \phi(\omega)} e^{i\omega t}= 
\\
=S(E)\times f(t)\nonumber
\end{eqnarray}
An arbitrary function of energy can obviously be added to the above
expression:
\begin{equation} 
F(E,t)=S_0(E)+ f(t)\times S(E)
\label{eq:lc}
\end{equation}
Thus, the light curves at different energies are related by a
linear transformation. 

\begin{figure}
\includegraphics[width=0.5\textwidth]{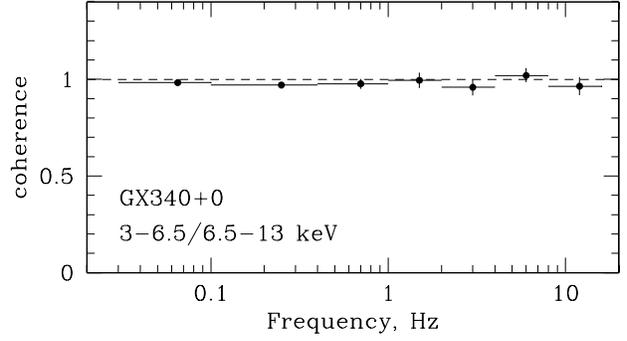}
\caption{GX340+0: Coherence between the light curves in the 3--6.5
  and 6.5--13 keV energy bands as function of frequency. No correction
  for the dead time effects has been made.  
\label{fig:coh_gx340}}
\end{figure}

From Eq.~(\ref{eq:lc})  it  follows that the coherence
of the signals in any two energy bands is exactly unity, as
they are related by a linear transformation. This prediction is in
a good agreement with observations (Fig.~\ref{fig:coh_gx340}).

\section{Boundary layer and accretion disk emission}
\label{sec:disk_bl}

As follows from eq.(\ref{eq:lc}), two components can be distinguished
in the GX340-0 emission.
The term $S_0(E)$ in Eq.~(\ref{eq:lc})  is the constant
(non-variable) part of the source emission
spectrum and $f(t)$ represents flux variations of the second, variable 
component.\footnote
{The possibility of small variations of the spectral parameter
(e.g. temperature, optical depth etc.) is discussed by
\citet{gilfanov03} and is shown to be inconsistent with observations. 
} 
The spectrum $S(E)$ of the variable component does
not change in the course of flux variations and equals the frequency
resolved spectrum, i.e. can be directly determined from observations.

There are two major components of  accretion onto  a slowly
rotating weakly magnetized neutron star -- (i) the Keplerian accretion
disk  and (ii) the boundary or spreading layer near
the surface of the neutron star, in which the accreting matter
decelerates to the spin frequency of the star and spreads over its
surface \citep{ss86, kluzniak, inogamov99, popham01}. 
These two geometrically distinct regions give comparable
contributions to the observed X-ray emission \citep{sibg00}.  
Recalling eq.(\ref{eq:lc}), it is plausible to assume, that the
variable part of the X-ray emission is associated with one of these 
components.   
In order to check this assumption and to  identify the variable
component we consider below theoretical expectations for the disk and
boundary layer spectra and compare them with the observed frequency
resolved spectra.

At sufficiently high values of $\dot{M}$ both BL and
disk are optically thick, as confirmed by the softness of the LMXB 
spectra. Simple arguments, taking into account the difference in
the  emitting areas  suggest that the spectrum of the boundary layer  
should be harder than that of the  accretion disk
\citep[e.g.][]{mitsuda84,greb}.
Due to complexity of the boundary/spreading layer problem the
theory has not advanced significantly beyond this
qualitative statement -- no models capable to directly predict its
spectrum  exist yet. 
Significantly better progress has been achieved in modeling  
spectra of accretion disks  \citep{ss73, shimura_takahara95,
ross96}. Relatively simple models of multicolor disk type which
account for the effects of Compton scattering with a 
simple color-to-effective temperature ratio  turned out to be
successful in describing the accretion disk spectra observed in the
high state of black hole systems \citep[e.g.][]{grad, gierlinski97}.

\begin{figure}
\includegraphics[width=\columnwidth]{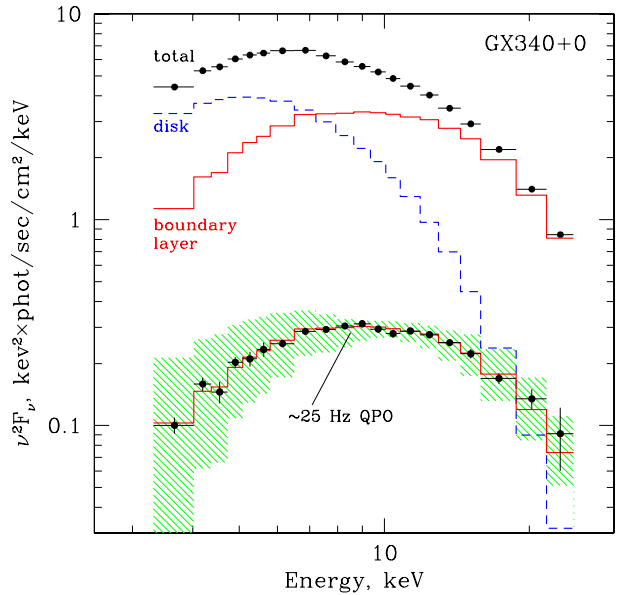}
\caption{The average and frequency resolved spectra of  GX340+0
(horizontal branch). 
The shaded area shows the plausible range of the boundary
layer spectra calculated as described in the section \ref{sec:disk_bl}.
The dashed (blue) histogram shows the best fit accretion disk
spectrum \citep{gilfanov03}. 
The upper solid (red) histogram shows the  boundary layer spectrum
computed as the difference between the (observed) total and
(predicted) accretion disk spectrum.  The lower solid histogram is the
same but scaled to the total energy flux of the frequency resolved
spectrum. 
\label{fig:disk_bl}}
\end{figure}

For this reason we chose to use a model of the disk emission as the
starting point. The BL spectrum is computed as a difference between 
the (observed) total spectrum and the (predicted) disk spectrum.
To estimate the plausible range of the BL spectra we
investigate the parameter space of the accretion disk model.
For the latter we adopt the general relativistic accretion disk
model by \citet{grad} (the ``grad model'' in XSPEC). 
The parameters of the model are: the source distance $D$,
mass of the central object $M_{\rm NS}$, disk inclination 
angle $i$, the mass accretion 
rate $\dot{M}$ and the color-to-effective temperature ratio 
$f=T_{\rm col}/T_{\rm eff}$. 
With this approach we can predict the disk and the boundary layer
spectra based on the observed X-ray flux and spectrum and very generic
system parameters, such as neutron star spin frequency, the source
distance etc.  The procedure is described in detail in
\citet{gilfanov03}. 
The obtained range of the BL spectra is shown in
Fig.~\ref{fig:disk_bl} as the shaded area. 
The similarity of the predicted BL spectrum and of the
observed frequency resolved spectrum is obvious. 
On the other hand the disk spectrum is significantly softer and is
inconsistent with the  frequency resolved spectrum. 

\citet{gilfanov03} conducted similar  investigation of an atoll
source 4U1608-52 having $\sim 10$ times lower luminosity. They
considered frequency resolved spectra at different frequencies,
including the kHz QPOs and showed that it demonstrates behavior
identical to GX340-0.  

Based on these results we conclude that the X-ray variability on
$\sim$ second -- millisecond time scales is related to variations of
the luminosity of the boundary layer. The shape of its spectrum
remains nearly constant in the course 
of these variations and equals the frequency resolved spectrum,
i.e. can be directly obtained from the observations. 
This can be used to separate the boundary layer and the accretion disk 
contribution to the total spectrum and permits to check quantitatively
the predictions the accretion disk and boundary layer models. 
It also opens the  possibility to measure relative contributions of
these two components of the accretion flow to the total observed X-ray 
emission.

\section{Boundary layer spectrum}
\label{sec:bl_spectrum}

\begin{figure}
\includegraphics[width=\columnwidth]{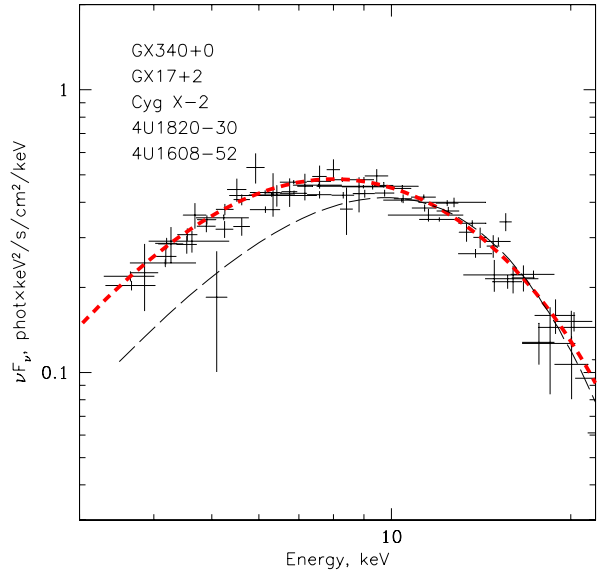}
\caption{Fourier-frequency resolved spectra ($\approx$boundary layer
spectra) of 5 Z- and atoll sources \citep[from][]{mikej05}. For
4U1608-52 the frequency resolved spectrum of the lower kHz QPO is
shown. All spectra were corrected for the interstellar absorption. The
thick short-dashed line shows the best fit Comptonization model with
$kT_s=1.5$, $kT_e=3.3$ keV, $\tau=5$.  
The thin long-dashed line shows the blackbody spectrum with temperature
$kT_{\rm bb}=2.4$ keV.} 
\label{freq_spectra}
\end{figure}

Based on the assumption that the frequency resolved spectra in
luminous LMXBs adequately represent the boundary layer spectrum, 
we compare several LMXBs -- atoll sources 4U1608-52 and
4U1820-30 and Z-sources Cyg X-2 and GX 17+2.
Their frequency resolved spectra   at  frequencies
$f\ga$ few Hz are shown in Fig.\ref{freq_spectra}.  
As before, for Z sources we used only data on the horizontal branch of
the color-color diagram, where the amplitude of variability at these 
frequencies is maximal. To facilitate comparison, the
normalizations of all spectra were adjusted  to match that of GX340+0.  
Similarity of the spectra is remarkable, especially considering
significant difference in the average spectra and a factor of $\sim
10-20$  spread in the luminosity between atoll  and Z-sources ($\sim
0.1 \dot{M}_{\rm Edd}$ and $\sim \dot{M}_{\rm Edd}$
correspondingly). 

The independence of the spectrum of the boundary layer on the
luminosity lends support to the theoretical predictions by
\citet{inogamov99} that the boundary layer is radiation pressure
supported, i.e. radiates at the local Eddington flux limit. 
In this model the luminosity of the spreading layer on the surface of
the neutron star changes due to variations of its area, rather than
surface emissivity of the unit area.
If this picture is correct, the parameters of the BL emission can be
used to determine the value of the Eddington flux limit on the surface
of the neutron star. As the Eddington flux limit is uniquely
determined by the neutron star surface gravity and the atmospheric
chemical composition, the  neutron star mass and radius can be
constrained \citep{mikej05}.

The similarity of the spectral shape of the BL spectrum in different
sources (Fig.\ref{freq_spectra}) indicates that there is no
significant spread in the values of the mass and radius among LMXBs,
in particular, that the surface gravity in atoll and Z-sources is 
similar. 
It also shows that there are no significant differences caused by
variations in the atmospheric chemical abundances between sources. 
In particular  we did not find statistically significant difference
between ultra-compact compact binary 4U1820-30 and other sources.

The shape of the frequency resolved ($\approx$ boundary layer)
spectrum can be adequately described by the saturated  
Comptonization. For the sake of comparison with other results
and for convenient parameterization of the  BL spectrum we used the
Comptonization model of Titarchuk (1994). The best fit parameters of
the model fitted in the 3-20 keV range simultaneously to all five
spectra shown in the Fig.\ref{freq_spectra} are:
temperature of seed photons $kT_s=1.5\pm 0.1$, temperature of
electrons $kT_e=3.3\pm0.4$ and the optical depth $\tau=5\pm1$ for slab
geometry. The best fit model is shown by the thick dotted line on
Fig.\ref{freq_spectra}. The temperature of the black body spectrum 
describing the shape of the  cutoff in the observed spectrum at
energies $>$13 keV is $kT_{\rm bb}=2.4\pm0.1$ keV (thin dashed line on 
Fig.\ref{freq_spectra}).

The fact that kHz QPO show the same behavior as other components
of the aperiodic variability indicates, that they have the same origin,
i.e. are caused by the variations of the luminosity of the boundary
layer. Although the kHz ``clock'' can be in the disk or due to it's
interaction with the neutron star, the actual modulation of the X-ray
flux  occurs on the neutron star surface.

\subsection{BL spectrum on the normal branch}

\begin{figure}
\includegraphics[width=0.5\textwidth, clip]{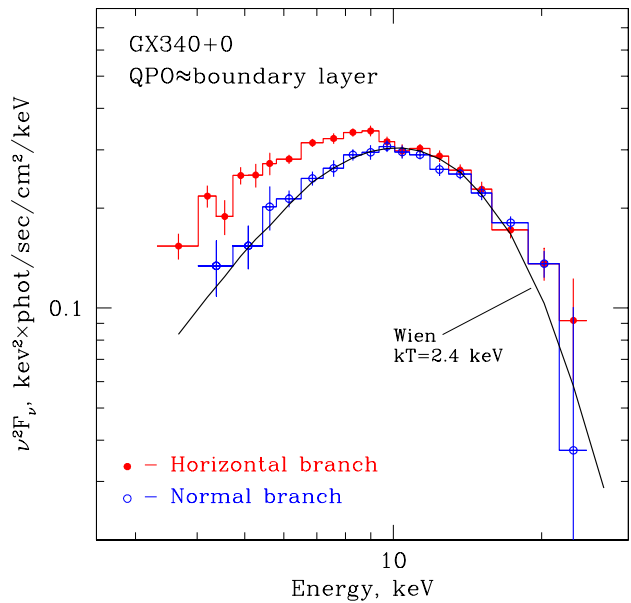}
\caption{The absorption corrected frequency resolved spectra of QPO
($\approx$ boundary layer emission) in GX340 on the horizontal branch,
(lower $\dot{M}$) and upper half of the normal branch
(higher $\dot{M}$). The horizontal branch data is same as in
Fig.~\ref{fig:freqres_gx340}--\ref{fig:disk_bl}.
The solid line shows  Wien spectrum with $kT=2.4$ keV.  
\label{fig:bl_alongz}}
\end{figure}

Further along the Z-track of GX340+0, on  normal and flaring branches,
the fractional rms of the X-ray variability decreases significantly,
by a factor of $\sim 5-10$. Nevertheless, the statistics is sufficient
to place meaningful constrains on the first half of the normal branch.  
The data indicates that the behavior of the  frequency resolved 
spectra does not change its character -- at sufficiently high frequency, 
$f\ga 1$ Hz, their shape does not depend upon the Fourier frequency
and is significantly harder than the average spectrum and expected
spectrum of the accretion disk.  
Therefore, we can conclude that frequency resolved spectra are
representative of the spectrum of the boundary layer.  
Fit to the frequency resolved spectrum by Comptonization model
requires infinitely large values of the  Comptonization parameter.
Correspondingly, it can be described by Wien or blackbody spectrum
(they are close to each other $E\ga 3$ keV range)  with the best fit
temperature of $kT\approx 2.4$ keV.  
The  frequency resolved spectra ($\approx$boundary layer spectra) of
GX340-0 on the normal and horizontal branch are  compared in
Fig.~\ref{fig:bl_alongz}. Thus, with increase of the mass accretion
rate up to a value close to critical Eddington rate the boundary layer
spectrum in the 3--20 keV energy  range approaches a Wien spectrum.

It would be interesting to follow up on these results  and to consider
the change of the BL spectrum  from the horizontal to normal branch in
other Z-sources.  
Unfortunately, in other four Z-sources from our sample the variability
level on the normal branch is insufficient to obtain frequency
resolved spectra with reasonable signal-to-noise ratio.

\section{Nature of the  Z-track}

Knowledge of the shape of the boundary layer spectrum allows us
to resolve the degeneracy caused by the similarity of the accretion
disk and boundary layer spectra,  which hindered many previous LMXB
studies. As demonstrated by \citet{gilfanov03}, the  spectra of atoll
and Z-sources can be adequately described by the sum of the
(renormalized) frequency resolved spectrum, representing the boundary
layer component,  and of the accretion disk emission
(Fig.~\ref{fig:disk_bl}).  
The spectrum of the latter is well described by the general
relativistic accretion disk model.  The best fit values  of the mass
accretion rate are consistent with those inferred from the observed
X-ray flux and accretion efficiency  appropriate for a 1.4$M_{\sun}$
neutron star with spin frequency of $\sim 500$ Hz.  
The agreement is especially remarkable, as the luminosity and mass
accretion rate in atoll and Z-sources differ by the factor of $\sim
10$.

We further exploit this approach and study the behavior of Z-sources 
in the color-color diagram in an attempt to relate the motion along
the Z-track to changes of the physically meaningful parameters.  
We consider spectra integrated over 128-sec time intervals. As
above, these spectra are fitted with a  model, consisting of the
boundary layer and the accretion disk components.   
The shape of the boundary layer spectrum was 
approximated by the $comptt$ model with parameters from
the section \ref{sec:bl_spectrum}. For the accretion disk spectrum
we adopt the multicolor disk model ($diskbb$ model in XSPEC) or
general relativistic accretion disk model ($grad$). 
The model adequately describes observed spectra on the normal and
horizontal branch. Further details of the analysis method, limitations
of the model and discussion of results are presented in
\citet{mikej05}.

\begin{figure}
\includegraphics[width=\columnwidth]{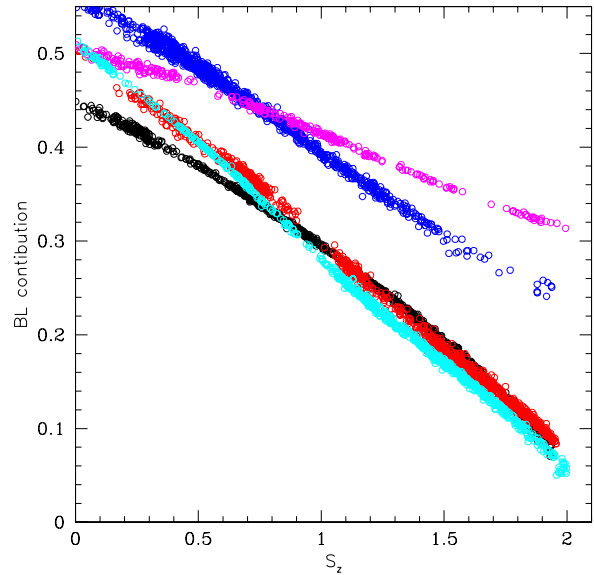}
\caption{Dependence of the boundary layer contribution to the total X-ray
emission of Z-sources as a function of the position on the Z-diagram}
\label{blcontr}
\end{figure}

\subsection{BL fraction}

The  dependences of the BL contribution to the total
X-ray emission  on the position on the Z-track are plotted
in Fig.\ref{blcontr}.  The coordinate along the Z-track was defined to
be proportional to the hard color with the reference points
$S_Z=1,2$ corresponding to the turning points of the ``Z'', as is 
commonly  used for such plots.
Statistical uncertainties in the values of the BL fraction
are small and can be neglected, as confirmed by the dispersion of  the
points in Fig.\ref{blcontr}. 
More important are the systematic ones associated with the
imprecise knowledge of the shape of the BL spectrum and its possible 
variations along the Z-track. As demonstrated in \citet{mikej05} their
amplitude  does not exceed $\sim 0.05-0.1$ in the  units of
Fig.~\ref{blcontr}. 

The Fig.\ref{blcontr} suggests that the boundary
layer fraction decreases along the Z-track and it is smaller on the
normal branch  than on the horizontal branch. As discussed in
\citet{mikej05}, this conclusion is rather robust, as long as the
assumption regarding the constancy of the boundary layer spectrum is
approximately correct. 
As the variability at $f\ga 1$ Hz is primarily associated with the
boundary layer emission, the decrease of the boundary layer fraction
along the Z-track also explains well-known decrease of the level of
aperiodic and quasi-periodic variability.

Although no simple physical interpretation of the observed behavior
can be offered, we mention several possibilities. One of these is that
the  general structure of the accretion flow does not change
significantly  and $\sim 50\%$ of the energy is always released on, or
very close to the neutron star surface. The apparent decrease of the
boundary layer  fraction on the normal branch is a result of its
geometrical obscuration by, for example, the geometrically thickened
accretion disk. 
An alternative possibility is that at high values of the mass
accretion rate  $\dot{M}\sim\dot{M}_{\rm Edd}$ a significant 
modification of the accretion flow structure occurs and its
division into two geometrically distinct parts -- boundary layer and
accretion disk, becomes inapplicable.
Namely, due to non-negligible pressure effects the deceleration of the
orbital motion of the accreting  matter from Keplerian frequency to
the neutron star spin frequency would take place in a geometrically
extended region with the  
radial extend of $\Delta R\sim R_{\rm NS}$. 
In this case, the  observed decrease of the boundary layer fraction
could reflect actual decrease of the fraction of the energy released
on the neutron star surface with the rest of the energy being released
in the extended transition region.

\begin{figure}
\includegraphics[width=\columnwidth]{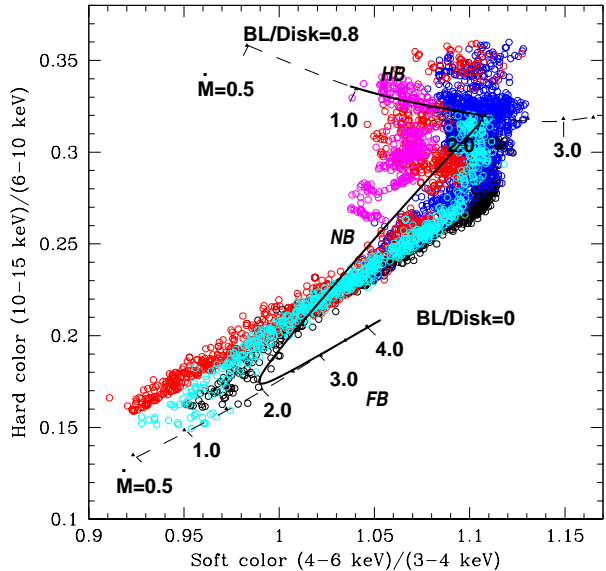}
\caption{The horizontal and normal branches in the color-color diagram
of several Z-sources (Cyg X-2, GX340+0, Sco X-1, GX 5-1 and GX 17+2).
Overlayed on the data are the tracks predicted by the model (section
\ref{sec:zdiagram}). The two thick dashed lines show evolution of the
colors with change of the accretion rate in the disk for two different
values of the BL fraction: 44\% (upper) and zero (lower)
The values of the  mass accretion rates $\dot{M}$ of the disk
component are marked in units of $10^{18}$ g/s.
The thick solid line shows the Z track with transition at 
$\dot{M}\sim 2\cdot 10^{18}$ g/s.}
\label{z_model}
\end{figure}

\subsection{Shape of Z-diagram}
\label{sec:zdiagram}

Motivated by these results we use the two-component spectral model
to explain the shape of the Z-track on the color-color diagram.  
For this purpose, the accretion disk component is modeled 
by the general relativistic accretion disk model of 
Ebisawa et al. (1991) (the {\em grad} model in  XSPEC) which
explicitly includes dependence on the mass accretion rate.   
Using this spectral model we can calculate the position
in the color-color diagram as a function of the mass accretion rate
and the boundary layer fraction.  
This is an attempt to understand  general tendencies
of the Z-diagram, rather than to construct its precise quantitative
description.

The results are presented in Fig.\ref{z_model}. The two dashed
curves in the figure show the evolution of spectral colors with 
the increase of $\dot{M}$ for two values of the boundary
layer fraction, BL/disk=0.8 and BL fraction of zero.
The evolution of colors corresponding to a change of the BL
contribution  from $F_{\rm BL}/F_{\rm disk}=0.8$ to zero 
at $\dot{M}\sim2\times 10^{18}$ g/s is shown by the thick solid line. 

The general shape of the Z-track can be reproduced in the model as a
result of variation of two parameters -- the mass accretion rate and
BL fraction.   
The mass accretion rate increases along the Z-track. The Z-shape of 
the track is defined by the variation of the BL fraction, which
decreases along the normal branch from the value of $\sim 50\%$
expected in the ``standard'' theories to a small number of the order
of $\sim$zero at the end of the normal branch.
The exact value of  $\dot{M}$, corresponding to the
transition from the horizontal to the normal branch depends on the 
disk model parameters -- the binary system inclination, 
the mass of the neutron star and the spectral hardening factor.
For our choice of parameters, it equals $\dot{M}\sim 2\cdot10^{18}$
g/sec, i.e. is of the order of the Eddington critical value for a 
$1.4M_\odot$ neutron star.

\section{Summary}

\begin{enumerate}

\item
The X-ray variability in luminous LMXBs on the short timescales,
$f\ga 1$ Hz, is caused by variations of the luminosity of the boundary
layer. The accretion disk emission is significantly less variable at
these frequencies. 
The BL spectrum remains nearly constant in the course of luminosity
variations and its shape equals the frequency resolved spectrum,
i.e. can be directly derived from the timing data
(Fig.\ref{fig:disk_bl}).  

\item
In the investigated range of the mass accretion rate $\dot{M}\sim
(0.1-1)\dot{M}_{\rm Edd}$, the boundary layer spectrum  depends weakly
on $\dot{M}$.  Its shape is remarkably similar in atoll and Z-sources
(Fig.~\ref{freq_spectra}), despite an order of magnitude difference  in
the mass accretion rate.
Data indicates that in the limit of high 
$\dot{M}\sim\dot{M}_{\rm Edd}$, 
the boundary layer spectrum can be described by Wien spectrum with
$kT\approx 2.4$ keV (Fig.~\ref{fig:bl_alongz}).  
At lower values of $\dot{M}$  
the spectra are better described by model of saturated Comptonization
with electron temperature of $\sim 2-4$ keV and Comptonization
parameter $y\sim 1$. 
Weak dependence of the BL spectrum  on the global value of $\dot{M}$
lends support to the  theoretical suggestion by \citet{inogamov99}
that the  boundary layer is radiation pressure supported. 

\item
The  kHz QPOs appear to have the same origin as aperiodic and
quasiperiodic variability at lower  
frequencies. The msec flux modulations originate on the surface of the
neutron star although the kHz ``clock''  might reside in the disk or
be determined by the  disk -- neutron star interaction.

\item
We  attempt to relate the motion of Z-sources along the Z-track to
changes in the values of the physically meaningful parameters.
Our results  suggest that the contribution of the boundary
layer  component to the observed emission decreases along the Z-track
from the conventional value of $\sim 50\%$ on the horizontal branch to
a rather small number at the end of the normal branch
(Fig.\ref{blcontr},\ref{z_model}).  
The main difference of our approach from previous attempts is in the
a priori knowledge of the shape of the boundary layer spectrum. This
allowed us to avoid ambiguity of the spectral decomposition into
boundary layer and disk components.

\end{enumerate}

\begin{acknowledgements}
This research has made use of data obtained through the High Energy
Astrophysics Science Archive Research Center Online Service, provided
by the NASA/Goddard Space Flight Center.
\end{acknowledgements}

\end{document}